\documentclass[aps,twocolumn,nofootinbib,preprintnumbers,superscriptaddress]{revtex4-2}

\usepackage{color}
\usepackage{bbm}
\usepackage{tikz}
\usepackage{braket}
\usepackage{comment}
\usepackage[
pagebackref=false,
colorlinks=true,
linkcolor=blue,
urlcolor=blue,
filecolor=black,
citecolor=red,
pdfstartview=FitV,
pdftitle={},
pdfauthor={},
pdfsubject={},
pdfkeywords={},
pdfpagemode=None,
bookmarksopen=true
]{hyperref}

\usepackage[normalem]{ulem}
\usepackage{amsmath, amssymb, bm}
\usepackage{enumerate}
\usepackage{amsfonts}
\usepackage{epsfig}
\usepackage{mathbbol}
\usepackage{mathrsfs}
\definecolor{darkgreen}{rgb}{0.0, 0.5, 0.0}
\usepackage{subcaption}

\usepackage{braket}
\usepackage{bm}
\newcommand{\be}{\begin{equation}}
\newcommand{\ee}{\end{equation}}
\newcommand{\f}{\frac}
\newcommand{\s}{\sqrt}

\def\ba#1\ea{\begin{align}#1\end{align}}

\usepackage{xcolor}
\begin{document}

\title{Lindbladian dynamics of the Sachdev-Ye-Kitaev model}

\author{Anish Kulkarni}
\affiliation{Department of Physics, Princeton University, Princeton, New Jersey, 08540, USA}

\author{Tokiro Numasawa}
\affiliation{
Institute for Solid State Physics, University of Tokyo, Kashiwa 277-8581, Japan}

\author{Shinsei Ryu}
\affiliation{Department of Physics, Princeton University, Princeton, New Jersey, 08540, USA}


\begin{abstract}
We study the Lindbladian dynamics of the Sachdev-Ye-Kitaev (SYK) model, where the SYK model is coupled to Markovian reservoirs
with jump operators that are either linear or quadratic in the
Majorana fermion operators. Here, the linear jump operators are non-random while the quadratic jump operators are sampled from a Gaussian distribution. In the limit of large $N$, where $N$ is the number of Majorana fermion operators, and also in the limit of large $N$ and $M$, where $M$ is the number of jump operators, the SYK Lindbladians are analytically tractable, and we obtain their stationary Green's functions,
from which we can read off the decay rate. For finite $N$, we also study the distribution of the eigenvalues of the SYK Lindbladians.
\end{abstract}

\maketitle

\section{Introduction}
While quantum dynamics is often modeled by an idealized unitary time evolution, non-unitary time evolutions are ubiquitous and relevant since experimental systems are never completely isolated. Non-unitarity may arise in many different forms, such as dissipation, gain/loss, decoherence, measurements, and so on. Understanding and controlling these effects are of both fundamental and practical importance. Furthermore, non-unitarity may give rise to rich behaviors that do not have counterparts in systems governed by unitary time evolution. Our understanding of possible universal behaviors in open quantum systems, however, is still limited, particularly in the context of many-body quantum systems and quantum field theory.

In this paper, we study tractable many-body quantum systems
with Lindbladian dynamics, aiming to deepen our understanding 
of open many-body quantum systems. The models we study 
consist of the SYK Hamiltonian, i.e., fermionic quantum many-body Hamiltonian with all-to-all interactions
\cite{1993PhRvL..70.3339S, KitaevKITP} 
and jump operators that we will describe momentarily.
Specifically,
we consider a set of Majorana fermion operators,
$
\{\psi_i, \psi_j\}
=\delta_{ij}
$,
$
\psi^{\dag}_i = \psi_i
$,
and the associated Fock space
where $i=1,\ldots, N$.
The Lindbladian $\mathcal{L}$
of our interest,
which generates
the dynamics
$d\rho/dt = \mathcal{L}(\rho)$,
is given by
\begin{align}
  \mathcal{L}
  (\rho)
  &=
    - i [H_{{\rm SYK}}, \rho]
    +
    \sum_{\alpha}
    \left[
    L^{\ }_{\alpha}
    \rho
    L^{\dag}_{\alpha}
    -
    \frac{1}{2}
    \{
    L^{\dag}_{\alpha}L^{\ }_{\alpha},
    \rho
    \}
    \right].
\end{align}
Here, the Hamiltonian part 
is given by 
the SYK (SYK$_q$) Hamiltonian
with $q$-body interaction,
\begin{align}
  H_{{\rm SYK}} &=
  i^{q/2}
      \sum_{i_1<\cdots < i_q}
      J_{i_1\cdots i_q}
      \psi_{i_1} \cdots \psi_{i_q},
\end{align}
where $J_{i_1\cdots i_q}$
are random couplings 
drawn from the Gaussian distribution.
As for the  jump operators
$\{L_{\alpha}\}$,
we consider the following two cases.
(i)
First,
we consider $N$
jump operators that are linear in
fermion operators, 
\be
\label{non-random jump-introduction}
L^i = \s{\mu}\psi^i,
\quad
i=1,\cdots, N.
\ee
Here, $\mu$ is a 
nonrandom real parameter.
(ii)
In the second example, 
we consider 
$M$
quadratic jump operators,
\begin{align}
L^a &= \sum_{1\leq i<j\leq N}
K^a_{ij} \psi_i \psi_j, \quad K^a_{ij} \in \mathbb{C},
\quad
a=1,\cdots, M
\end{align}
where all $K^a_{ij}$ are independent complex Gaussian distributed random variables with mean and variance given by
\begin{equation}
	\langle
	K^a_{ij}\rangle = 0, \quad 
	\langle
	|K^a_{ij}|^2\rangle = \frac{K^2}{N^2} \quad \forall\ i,j,a \quad \text{(no sum)}.
\end{equation}
It is also possible to consider more generic jump operators
that consist of $p$ Majorana operators. 
For more details on these models, see later sections. 
As we will show, these models can be analytically studied in the limit $N\to \infty$ in the first example, 
and in the limit $N,M\to \infty$, with fixed $R=M/N$, in the second example.

The SYK model and its variants have provided various tractable toy models
for many-body problems, e.g., 
the butterfly effect, quantum information scrambling, 
quantum entanglement, 
non-Fermi liquids, etc., 
and have been extensively studied recently
\cite{Polchinski:2016xgd,
2016PhRvD..94j6002M,
Gu:2016oyy,
2017PhRvL.119u6601S,
Altland:2019lne,
2019JPhA...52.3001R,
2021arXiv210905037C}.
Some of these models admit holographic dual descriptions.
We note that nonunitary time evolution of various kinds 
in SYK-type models has also been studied recently.
See, for example,
\cite{
2020arXiv200811955L,
2021PhRvD.103d6014G,
2021arXiv210206630G,
2021arXiv210404088Z,
2021arXiv210408270J,
2021arXiv210811973J,
2021arXiv210806784C,
2021arXiv210903268Z,
2021PhRvB.103f4309X,
2021JHEP...06..156S,
2021arXiv211003444G,
2021arXiv211208373A}.
Our study using the Lindbladian dynamics is different from and complementary to these previous works. 
The effects of dissipation in the SYK models have also 
been studied within unitary dynamics by including the heat bath degrees of freedom explicitly.
There are two-coupled variants of SYK models, where one of the copies
can be considered a bath. See, for example,
\cite{2018arXiv180400491M,
2019arXiv191203276M,
2019PhRvX...9b1043K}.
At more technical levels, there are other
Hermitian SYK type models (supersymmetric SYK and Wishert SYK models) that have some resemblance 
to our SYK Lindbladian model(s)
\cite{2017PhRvD..95f9904F,
2017PhRvB..95t5105B,
2021arXiv210407647S}.

In this work, we will study the properties of the above SYK
Lindbladians by computing Green's functions
in the large-$N$ (and large-$M$) limit. This allows us to extract, for example, the dominant decay rate (the spectral gap of the Lindbladians).
We also study spectral properties of finite-$N$ versions of the SYK Lindbladians 
by diagonalizing them numerically.
The spectrum is complex in general and distributed non-trivially over the complex plane. 
Because of the all-to-all nature of 
interactions 
(and jump operators in the second case),
the spectrum can be naturally compared with known 
behaviors in
random Lindbladians studied by using techniques from 
Random Matrix Theory
\cite{2019PhRvL.123n0403D,
2019PhRvL.123w4103C,
2020PhRvL.124j0604W,
2019JPhA...52V5302C,
2020JPhA...53D5303S,
2020PhRvB.102m4310S,
2020PhRvX..10b1019S,
2021arXiv211013158L,
2021PhRvE.104c4118T}. 
For the case of random quadratic jump operators,
the spectrum crosses over from elliptic disk-shaped 
to ``lemon-shaped" distribution by increasing the relative strength 
between the SYK interaction and the jump operators, $K/J$.
The latter is a ubiquitous behavior in the strong dissipation regime
of random Lindbladians
\cite{2019PhRvL.123n0403D, 2019PhRvL.123w4103C, 2020JPhA...53D5303S}. 
We also observe clustering of eigenvalues by controlling the ratio $R=M/N$
\cite{2019PhRvL.123w4103C,
2020PhRvL.124j0604W, 2021arXiv211013158L}.

\section{
The operator-state isomorphism
and 
the Schwinger-Keldysh path integral}
\label{The operator-state isomorphism and  the Schwinger-Keldysh path integral}

We are interested in 
the spectral properties
of the Lindbladians,
and various correlation functions
(mostly in the large-$N$ limit).  
To this end, 
we will set up a 
Schwinger-Keldysh type
path-integral approach
that involves
two copies of 
path-integral variables
\cite{2016RPPh...79i6001S, 2017arXiv170607803E, kamenev_2011}. 
We do so by first invoking 
the state-operator map
(the Choi-Jamio\l kowski isomorphism)
to ``vectorize" the Lindbladian.
This allows us to think of  operators
(the density matrix in particular)
as a state in the  doubled Hilbert space,
$\mathcal{H}\otimes \mathcal{H}^*
\equiv 
\mathcal{H}_+\otimes \mathcal{H}_-
$,
and 
$\mathcal{L}$ 
as an operator acting 
on the doubled Hilbert space.
The first step in the state-operator map is  to consider 
a maximally entangled state 
$|\mathbb{I}\rangle$
in 
$\mathcal{H}_+\otimes \mathcal{H}_-$.
This state should have a property 
that it maps or ``reflects"
all operators on the first Hilbert space 
to  corresponding ones 
in the second Hilbert space (and 
vice versa):
$
O_+ | \mathbb{I}\rangle = O'_- |\mathbb{I}\rangle
$,
where $O_{+}$ and $O'_{-}$ are some operators 
acting on the first and second Hilbert spaces.
In particular, we require
\begin{align}
\psi^i_+ |\mathbb{I}\rangle = -i \psi^i_- |
\mathbb{I}\rangle.
\label{reflection-condition}
\end{align}
The factor of $-i$ originates from the Fermi statistics:
reflecting twice gives a $2\pi$ rotation
under which the fermion operators pick up $-1$.
With $|\mathbb{I}\rangle$ in hand, we can map
an operator, the density matrix $\rho$, say, 
to the corresponding state on 
$\mathcal{H}_+\otimes \mathcal{H}_-$
as
\begin{align}
\rho \longrightarrow 
|\rho\rangle \equiv
\rho_+ |\mathbb{I}\rangle  \in 
\mathcal{H}_+\otimes \mathcal{H}_-.
\end{align}
Note that the identity operator 
$\mathbb{I}$, which can be thought of as 
an infinite temperature Gibbs state,
is mapped to $|\mathbb{I}\rangle$.
Similarly, the Lindbladian can be mapped 
to an operator acting on 
$\mathcal{H}_+\otimes \mathcal{H}_-$,
and
the 
Lindblad equation is now written as
$
d |\rho\rangle/dt = \mathcal{L}|\rho\rangle
$,
where we continue to use $\mathcal{L}$ to represent 
the mapped operator. 
The explicit form of $\mathcal{L}$
for our models is given in equations \eqref{non-rand lindbrad} and \eqref{Random lindblad}.
The state $\ket{\mathbb{I}}$ is annihilated by $\mathcal{L}$,
$
\mathcal{L} \ket{\mathbb{I}} = 0
$,
as the infinite temperature
state is stationary 
with respect to 
any Lindbladian.

With the operator-state map,
for example, 
the ``partition function"
can be expressed as
$
\mathrm{Tr}\, [\rho(t)]
=
\langle \mathbb{I}| \rho(t)\rangle
=
\langle \mathbb{I}| e^{t \mathcal{L}} | \rho_0\rangle
=
\langle \mathbb{I}| \rho_0\rangle = 1
$
where $|\rho_0\rangle$ is an initial condition
and we noted $\langle \mathbb{I}| e^{ t \mathcal{L}} = \langle \mathbb{I}|$.
Similarly,
the expectation 
value of an operator 
$A$ is given by
$
\mathrm{Tr}\, 
[\rho(t) A]
=
\langle \mathbb{I}| 
A_+ \otimes \mathbb{I}_-
e^{t \mathcal{L}} | \rho_0\rangle
$.
These quantities can be readily 
expressed 
in terms of
the coherent state path integral
over two copies of real fermionic fields,
$\psi^i_{\pm}(t)$,
as
\be
\label{SK path integral}
Z = 
\braket{\mathbb{I}|e^{t \mathcal{L}}|\rho_0} 
= \int \mathcal{D}\psi_+\mathcal{D}\psi_- e^{iS[\psi_+,\psi_-]},
\ee
i.e., the Schwinger-Keldysh formalism.

For the SYK type models discussed below, we will analyze 
the Schwinger-Keldysh 
path integral 
\eqref{SK path integral}
in the large-$N$ limit.
Furthermore, 
in this work, 
we will be interested
in stationary properties 
that may emerge 
in the late time limit.
In particular,
we will assume
in this limit
that
the memory 
of the initial state is lost, 
and the system relaxes 
into a stationary state
independent of the initial state.


\begin{widetext}
\section{Non-random linear jump operators}

In this section, we consider
the SYK model
in the presence of 
the jump operators
\be
\label{non-random jump}
L^i = \s{\mu}\psi^i,
\quad 
i=1,\cdots, N.
\ee
Here, we assume $\mu$ is a real parameter.
Following the procedure outlined in the previous section,
the Lindbladian acting on the
doubled Hilbert space $\mathcal{H}_+\otimes \mathcal{H}_-$
is given by
\be
\label{non-rand lindbrad}
\mathcal{L} =  -iH_{{\rm SYK}}^+\otimes \mathbb{I}_- + i  (-1)^\f{q}{2}\mathbb{I}_+\otimes 
H_{{\rm SYK}}^-    -i \mu \sum_i\psi_+^i\psi_-^i -\mu \f{N}{2}
\mathbb{I}_+ \otimes  \mathbb{I}_-.  
\ee

At least superficially, this model looks
similar to the two-coupled SYK model
discussed in \cite{2018arXiv180400491M}.
We however note various differences.
The first is the relative phase between the $H_{{\rm SYK}}^+$ and $H_{{\rm SYK}}^-$ terms.
For example, when $q=4$,
we have opposite signs for these terms. 
The relative sign between the terms is necessary
so that their sum is an isometry of $\ket{\mathbb{I}}$.
On the other hand, for the regular two-coupled SYK model,
these terms have the same sign,
and induces time evolution. 
Another difference is that the Hamiltonian terms
(the first two terms) are anti-Hermitian
whereas $ -i \mu \sum_i\psi_+^i\psi_-^i$ is Hermitian.
Overall, $\mathcal{L}$ is not anti-Hermitian 
($\mathcal{L}^{\dagger}\neq -\mathcal{L}$) and evolution is nonunitary.

\subsection{Path integral and large-$N$ effective action}
Using the formalism 
in the previous section,
we can study this model
using the Schwinger-Keldysh path
integral.
The action is given by
\ba
iS[\psi_+,\psi_-]
&= \int_{t_i}^{t_f} dt \Bigg[ -\f{1}{2} \sum_i \psi^i_+ \partial_t \psi^i_+-\f{1}{2} \sum_i \psi^i_- \partial_t \psi^i_- 
  - i^{q+1} \sum_{i_1 <\cdots< i_q} J_{i_1\cdots i_q} \psi_+^{i_1} \cdots \psi_+^{i_q} \notag \\
  &\qquad+ i^{q+1} \sum_{i_1 <\cdots< i_q} J_{i_1\cdots i_q} \psi_-^{i_1} \cdots \psi_-^{i_q} -i \mu \sum_i\psi_+^i(t) \psi_-^i(t) - \mu \f{N}{2} \int dt 1 
 \Bigg].
\ea
The action has to be 
supplemented with the 
proper boundary conditions
at $t=t_i, t_f$,
set by the initial 
($|\rho_0\rangle$)
and final ($|\mathbb{I}\rangle$) states.
When analyzing the stationary state, however, the boundary conditions 
are immaterial. 
This path integral can be studied in
the large $N$ limit as in the regular SYK model.
We introduce
two kinds of matrix collective fields,
$G_{\alpha\beta}(t_1,t_2)$
and
$\Sigma_{\alpha\beta}(t_1,t_2)$,
where $\alpha,\beta\in \{+,-\}$.
The effective action for the collective fields is  
\ba
S[G,\Sigma] &= -\f{i N}{2} 
\ln\det\, [-i (G_0^{-1} - \Sigma) ]  + \f{i^{q+1}J^2 N}{2 q} \int_{t_i}^{t_f} dt_1 dt_2  \sum_{\alpha \beta}s_{\alpha \beta}G_{\alpha\beta}(t_1,t_2)^{q} \notag \\
& \qquad + \f{iN}{2}\int_{t_i}^{t_f} dt_1 dt_2  \sum_{\alpha \beta} \Sigma_{\alpha\beta}(t_1,t_2) G_{\alpha \beta}(t_1,t_2) - i \f{\mu N}{2} \int_{t_i}^{t_f} dt [G_{+-}(t,t) - G_{-+}(t,t)] +i \mu \f{N}{2} \int dt,
\ea
where $s_{\alpha\beta}$ is given by 
\be
s_{++} = s_{--} = 1, \qquad s_{+-} = s_{-+} = -(-1)^{\f{q}{2}}.
\ee
In the saddle point approximation, the collective field $G_{\alpha \beta}$ is nothing but
the Green's functions of the fermion fields, 
\be
G_{\alpha\beta}(t_1,t_2) = -i \braket{T(\psi_\alpha(t_1)\psi_{\beta}(t_2))}.
\ee
The correlation functions  satisfy the symmetry relation $G_{\alpha\beta}(t_1,t_2) = -G_{\beta\alpha}(t_2,t_1) $.
The partition function of the system 
in terms of the corrective fields 
is 
$
Z = \int \mathcal{D}G_{\alpha\beta} \mathcal{D}\Sigma_{\alpha\beta} \exp\{i S[G,\Sigma]\}.
$
The large $N$ saddle point equation is 
\ba
  &
    \label{eq:KBequationNonRandom1}
    i\partial_{t_1} G_{\alpha \beta}(t_1,t_2) - \int dt_3 \sum_{\gamma = +, -}\Sigma_{\alpha\gamma}(t_1,t_3) G_{\gamma \beta}(t_3,t_2) = \delta _{\alpha\beta}\delta(t_1-t_2),
  \\
  &
  \label{eq:KBequationNonRandom2}
  \Sigma_{\alpha\beta}(t_1,t_2) = -i^{q}J^2 s_{\alpha\beta}G_{\alpha\beta}(t_1,t_2)^{q-1} + \mu\epsilon_{\alpha\beta} \delta(t_1-t_2).
\ea
\end{widetext}
%

\subsection{Stationary Green's functions}

\begin{figure}[t]
\begin{center}
\includegraphics[scale=0.9]{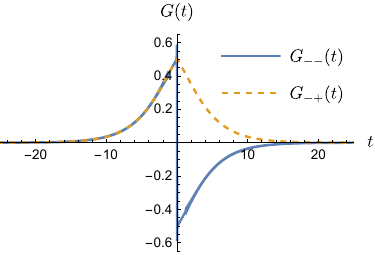}
\hspace{1cm}
\includegraphics[scale=0.9]{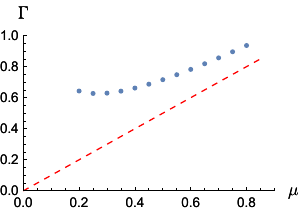}
\end{center}
\caption{
Top: The Green's functions
in the large-$N$ limit
for $q=4$, $J = 1$, and $\mu = 0.250$.
Bottom:
The decay rate $\Gamma$ of the correlation functions as a function of $\mu$. 
For large $\mu/J$,
the decay rate 
approaches $\Gamma=\mu$ (red dashed line).
}  
\label{fig:CorFunJ1Mu250}
\end{figure}

\paragraph{Large-$N$ limit with $q=4$}
The saddle point equation can be analyzed 
numerically, or by taking the large $q$ limit. 
We first take $q=4$
and solve the Kadanoff-Baym equations \eqref{eq:KBequationNonRandom1} and
\eqref{eq:KBequationNonRandom2} numerically.
We note that
assuming the memory of the initial state 
is lost in the long time limit, 
the time translation invariance is recovered
and 
the collective fields depend only on
$t_1-t_2 \equiv t$.
In Fig.\ \ref{fig:CorFunJ1Mu250}, 
we show an example of 
the numerical stationary solution
for $ J = 1$ and $\mu = 0.250$.
For large enough $\mu \gg J$, 
the system crosses over 
to the case of a dissipation-only model, 
where the correlation function
decays exponentially with the decay rate given $\Gamma$ approaching $\mu$, as shown in Fig.\ \ref{fig:CorFunJ1Mu250}. This is consistent with the spectrum at finite $N$, where an isolated cluster of eigenvalues forms at $-\mu$ when $\mu$ is large.
\paragraph{Large-$q$ limit}
In the large-$q$ limit, we expand the Green's function as 
\cite{2016PhRvD..94j6002M}
\ba
G_{\alpha\beta}(t_1,t_2) &= G^0_{\alpha\beta}(t_1,t_2) \Big( 1 + \f{1}{q} g_{\alpha\beta}(t_1,t_2) + \cdots \Big), 
\ea
The Kadanoff-Baym equation then reduces to the Liouville equation
\ba
\partial_{t_1}\partial_{t_2} g_{++}(t_1,t_2) &= - 2\mathcal{J}^2 e^{g_{++}(t_1,t_2)}, \notag \\
\partial_{t_1}\partial_{t_2} g_{+-}(t_1,t_2) &= - 2\mathcal{J}^2 e^{g_{+-}(t_1,t_2)} -2\hat{\mu}\delta(t_1-t_2).
\ea
Here we have defined $J^2 = \f{2 ^{q-1}\mathcal{J}^2}{ q}$ and $\mu = \f{\hat{\mu}}{q}$.
We impose the boundary conditions as 
\begin{align}
&
g_{++}(t,t) = 0, 
\nonumber \\
&
\lim_{t_2 \to t_1}\partial_{t_1}g_{+-}(t_1,t_2) = - \hat{\mu}, 
\nonumber \\
&
g_{++}(t_1,t_2) - g_{+-}(t_1,t_2) \to 0 \text{ as } t_1 \to \infty.  
\end{align}
We can then obtain a stationary solution as 
\begin{align}
&
e^{g_{++}(t)} = \f{\alpha^2 }{\mathcal{J}^2 \cosh^2 (\alpha |t| + \gamma )}, 
\nonumber \\
&
e^{g_{+-}(t)} = \f{\tilde{\alpha}^2 }{\mathcal{J}^2 \cosh^2 (\tilde{\alpha} |t| + \tilde{\gamma} )}.
\end{align}
To satisfy the boundary conditions, we impose 
\be
 \f{\alpha}{\mathcal{J} \cosh \gamma} = 1, 
 \quad \hat{\mu} = 2\tilde{\alpha} \tanh \tilde{\gamma},
 \quad \alpha = \tilde{\alpha}, 
 \quad \gamma = \tilde{\gamma}.
\ee
By solving these conditions, we obtain
\be
\alpha = \tilde{\alpha} = \mathcal{J} \s{\Big(\f{\hat{\mu}}{2\mathcal{J}}\Big)^2 + 1} , \quad \gamma = \tilde{\gamma} =  \text{arcsinh}\Big( \f{\hat{\mu}}{2\mathcal{J}} \Big).
\ee
From these, we see that the correlation functions behave
as $G(t) \sim e^{\f{g(t)}{q}}$ and decay exponentially.
We can read off
$\f{2 \alpha}{q} \equiv \Gamma$ as the decay rate.
This behavior also qualitatively agrees with the $\mu$ dependence
for the $q=4$ case above analyzed numerically.
Also, for large $q$,
we can confirm that
as $\mu \to 0$
(after taking the long-time limit),
the Green's function reduces to 
the infinite temperature thermal Green's function.
%
%
%
\subsection{Finite $N$ spectrum}
\begin{figure*}[t]
\begin{center}
    \begin{subfigure}{0.46 \textwidth}
        \caption{$\mu=0.1$}
        \includegraphics[width=\textwidth]{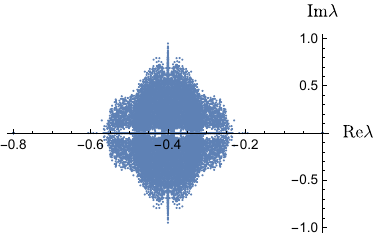}
    \end{subfigure}
    \begin{subfigure}{0.46 \textwidth}
        \caption{$\mu=0.3$}
        \includegraphics[width=\textwidth]{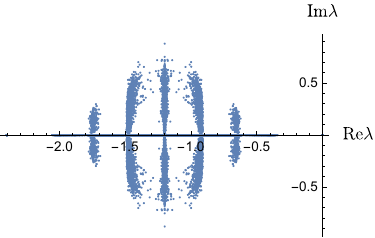}
    \end{subfigure}\\
    \vspace*{0.7cm}
    \begin{subfigure}{0.46 \textwidth}
        \caption{$\mu=0.5$}
        \includegraphics[width=\textwidth]{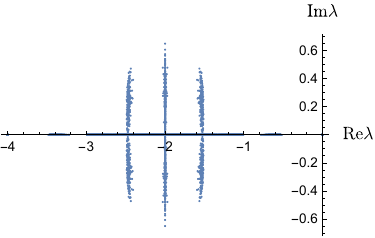}
    \end{subfigure}
    \begin{subfigure}{0.46 \textwidth}
        \caption{$\mu=0.9$}
        \includegraphics[width=\textwidth]{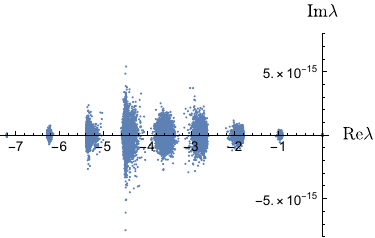}
    \end{subfigure}
\caption{Spectrum of the 
SYK Lindbladian operator $\mathcal{L}$ 
\eqref{non-rand lindbrad}
for $\mu=0.1, 0.3, 0.5$ and $0.9$
with $J=1$.}  
\label{fig:specN8}
\end{center}
\end{figure*}

We now turn 
to the spectral properties 
of the SYK Lindbladian
\eqref{non-rand lindbrad}.
The complex spectrum 
$\{\lambda_i\}$
of the SYK Lindbladian
\eqref{non-rand lindbrad}
can be studied 
by numerical exact diagonalization
for finite $N$. We set $N=8$ in our analysis below,
which means, 
including both copies
$\psi^{i}_+$ and $\psi^{i}_-$,
we have $2N=16$ flavors of 
Majorana fermion operators. Plotted in Fig.\ \ref{fig:specN8} are
the numerical spectra $\{\lambda_i\}$
for representative choices of $\mu$ 
(we set $J=1$). For each $\mu$, $100$ disorder realizations were collected.

For small $\mu$, there are many 
eigenvalues centered around 
$\text{Re}\, (\lambda) = -N\mu/2$.
As we increase $\mu$, vertical 
bands of eigenvalues
start forming
along the real axis. Each band is
located roughly along a line ${\rm Re}\,(\lambda) = -\mu n$ 
for $n = 1, \ldots, N$.
As we increase $\mu$ even further,
all the eigenvalues become close to real. This reminds us of a real-complex transition in some non-Hermitian systems \cite{2019PhRvL.123i0603H}.
Another effect of increasing $\mu$ is the formation of clusters around $\lambda=-n\mu$, with gaps in between. The cluster formation first occurs at the left and right edges of the spectrum, i.e. at small and large $n$, and then subsequently occurs at intermediate values of $n$. Similar band and cluster formation 
and hierarchy of relaxation times
were observed in 
Refs.\ \cite{2020PhRvL.124j0604W, 2021arXiv211013158L},
although we should note that these works
studied purely dissipative random
Lindbladians, while in our model the randomness enters
only in the Hamiltonian part.

In the weak dissipation regime, that is $\mu/J < 0.5$, the non-linearity in the decay rate and the band-formation in the spectrum indicates that there is a non-trivial competition between the dissipative and SYK interactions. 
%
%
\begin{widetext}
\section{Random quadratic jump operators}
\label{Many Random Jump Operators (Complex coupling)}
In this section we consider another open SYK system. Here we introduce $M$ two-body jump operators $L^a$ with random couplings. The Hamiltonian and the jump operators respectively are:
\begin{equation}
\label{random-quad-model}
	\begin{split}
		H_{\text{SYK}} &= i^{q/2} \sum_{1\leq i_1 < \cdots < i_q \leq N} J_{i_1\cdots i_q} \psi_{i_1} \cdots \psi_{i_q},
		\\
		L^a &= \sum_{1\leq i<j\leq N} K^a_{ij} \psi_i \psi_j, \quad K^a_{ij} \in \mathbb{C}, \quad a = 1, \cdots, M.
	\end{split}
\end{equation}
All $K^a_{ij}$ are independent identically distributed complex Gaussian random variables with mean and variance given by
\begin{equation}
\langle
	{K^a_{ij}}\rangle = 0, \quad 
	\langle
	{|K^a_{ij}|^2}\rangle = \frac{K^2}{N^2} \quad \forall i,j,a \quad \text{(no sum)}.
\end{equation}
The Lindbladian acting on the doubled Hilbert space $\mathcal{H}_+\otimes \mathcal{H}_-$
is given by
\begin{equation}
\label{Random lindblad}
    \begin{split}
        \mathcal{L} 
        &=  
        - iH_{{\rm SYK}}^+\otimes \mathbb{I}_- 
        + i (-1)^\f{q}{2}\mathbb{I}_+ \otimes H_{{\rm SYK}}^-
        - \sum_{a} L_{+}^{a} \otimes L^a_{-}{}^{\dagger}
        - \f{1}{2} \sum_{a} L^a_+{}^{\dagger}L^a_+ \otimes \mathbb{I}_- 
        - \f{1}{2} \mathbb{I}_+ \otimes\sum_{a} L^{a}_{-} L^a_{-}{}^{\dagger}\\
        &=
        - iH_{{\rm SYK}}^+\otimes \mathbb{I}_- 
        + i (-1)^\f{q}{2}\mathbb{I}_+ \otimes H_{{\rm SYK}}^- 
        + \sum_{a} \sum_{i<j} \sum_{k<l} K^a_{ij} \bar{K}^a_{kl} \left(          \psi^i_{+}\psi^j_{+}\psi^k_{-}\psi^l_{-}
        + \frac 12 \psi^k_{+}\psi^l_{+}\psi^i_{+}\psi^j_{+}
        + \frac 12 \psi^i_{-}\psi^j_{-} \psi^k_{-}\psi^l_{-} \right).
    \end{split}
\end{equation}

\subsection{Path integral and large-$N$ effective action}

Using the formalism in Section \ref{The operator-state isomorphism and  the Schwinger-Keldysh path integral}, we can obtain the Schwinger-Keldysh action for this model. Since we will be analyzing the stationary state, the initial state $\ket{\rho_0}$ is immaterial, and will be excluded from the path integral. We introduce complex Hubbard-Stratonovich (or auxiliary) fields $b^a_+(t)$ and $\ b^a_-(t)$ to make the dissipation term linear with respect to the jump operators. The resulting action is as follows:
\begin{equation}
    \begin{split}
        &iS[\psi_+,\psi_-, b^a_+, b^a_-, \bar b^a_+, \bar b^a_-] \\
        =& \int dt \Bigg[ 
        -\f{1}{2} \sum_i \psi^i_+ \partial_t \psi^i_+ 
        - \f{1}{2} \sum_i \psi^i_- \partial_t \psi^i_-
        - i^{q/2+1} \sum_{i_1 <\cdots< i_q} J_{i_1\cdots i_q} \psi_+^{i_1} \cdots \psi_+^{i_q}
        - (-i)^{q/2+1} \sum_{i_1 <\cdots< i_q} J_{i_1\cdots i_q} \psi_-^{i_1} \cdots \psi_-^{i_q}\\
        &\qquad \qquad - \frac 12 \sum_a 
        \begin{pmatrix}
                \bar{b}^a_+(t) & \bar{b}^a_-(t)
        \end{pmatrix}
        \begin{pmatrix}
                1 & 0 \\
                -2 & 1
        \end{pmatrix}
        \begin{pmatrix}
                b^a_+(t)\\
                b^a_-(t)
        \end{pmatrix}\\
        &\qquad \qquad + \frac 12 \sum_a \left( 
        \bar b^a_+(t)L^a_+(t) + 
        \bar b^a_-(t)L^a_-(t) + 
        b^a_+(t) \bar L^{a}_+(t) + b^a_-(t) \bar L^{a}_-(t) \right)
 \Bigg].
    \end{split}
\end{equation}
We then perform disorder averaging over the random couplings $J$ and $K$. Next, we introduce collective fields for both, the fermion fields and the auxiliary fields. We denote the fermion collective fields by $G_{\alpha\beta}$ and $\Sigma_{\alpha\beta}$, and the auxiliary collective fields by $G^b_{\alpha\beta}$ and $\Sigma^b_{\alpha\beta}$, where $\alpha,\beta = \pm$. Consider the limit $N,M \rightarrow \infty$ with constant $R=M/N$. In this limit, the Green's functions and self energies of the system are determined by the saddle point of the action. The saddle point equations are as follows:
\begin{equation}
	\begin{split}
		\Sigma^b_{\alpha\beta}(t_1,t_2) =& \frac{K^2}{4} G_{\alpha\beta}(t_1,t_2)^2,
		\\
		\mathbf{G}^b(t_1,t_2) =& \left[ 
		\begin{pmatrix}
			1 & 0 \\
			-2 & 1
		\end{pmatrix}\delta(t_1-t_2) - \mathbf{\Sigma}^b(t_1,t_2) \right]^{-1},
		\\
	\Sigma_{\alpha\beta}(t_1,t_2) =& -i^q J^2 s_{\alpha\beta} G_{\alpha\beta}(t_1, t_2)^{q-1} + \frac{K^2 R}{2} \left( G^b_{\alpha\beta} (t_1, t_2) + G^b_{\beta\alpha} (t_2, t_1) \right) G_{\alpha\beta} (t_1, t_2),
	\\
	\mathbf{G}(t_1,t_2) =& \left[ \mathbf{G}_0^{-1}(t_1,t_2) - \mathbf{\Sigma}(t_1,t_2) \right]^{-1}. 
	\end{split}
\end{equation}
The boldface fields are $2\times 2$ matrices with $\pm$ indices. The matrix inverses are with respect to this $2\times 2$ matrix multiplication as well as the time domain multiplication. 
Now let us apply the stationary state hypothesis to obtain the Schwinger-Dyson equations:
\begin{equation}
\label{random-model-SD-equations}
	\begin{split}
		\Sigma^b_{\alpha\beta}(t) =& \frac{K^2}{4} G_{\alpha\beta}(t)^2,
		\\
		\mathbf{G}^b(\omega) =& \left( 
		\begin{pmatrix}
			1 & 0 \\
			-2 & 1
		\end{pmatrix} - \mathbf{\Sigma}^b(\omega) \right)^{-1},
		\\
		\Sigma_{\alpha\beta}(t) =& -i^q J^2 s_{\alpha\beta} G_{\alpha\beta}(t)^{q-1} + \frac{K^2 R}{2}  \left( G^b_{\alpha\beta} (t) + G^b_{\beta\alpha} (-t) \right)G_{\alpha\beta} (t),
		\\
		\mathbf{G}(\omega) =& ( \mathbf{G}_0^{-1}(\omega) - \mathbf{\Sigma}(\omega) )^{-1}. 
	\end{split}
\end{equation}

\end{widetext}
\subsection{Stationary Green's functions}
\begin{figure}[t]
\centering
		\includegraphics[scale=1]{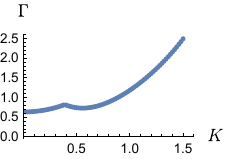}
		\hspace{0.1cm}
		\includegraphics[scale=1]{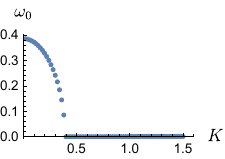}
	\caption{Decay rate $\Gamma$
	and frequency $\omega_0$
	of late time $G^R(t)$ for $J=1,R=2$.}
	\label{decay-and-oscillations}
\end{figure}

We solve \eqref{random-model-SD-equations} numerically for $q=4$ and various values of the parameters $J,K,$ and $R$. For all these solutions, the Green's functions of the Hubbard-Stratanovich fields are numerically consistent with the following trivial solution:
\begin{equation*}
    \mathbf{G}^b(t) = 
    \begin{pmatrix}
		1 & 0 \\
		2 & 1
	\end{pmatrix}
	\delta(t).
\end{equation*}
In all cases, the fermion Green's functions decay exponentially at late times. For small dissipation strength, the Green's functions oscillate as they decay. To characterize these oscillation we try to fit the retarded Green's function $G^R(t) = -i \Theta(t)[G_{+-}(t) - G_{-+}(t)]$ with the following ansatz at late times.
\begin{equation}
    G^R(t) \approx A e^{-\Gamma t}\sin(\omega_0 t + \phi).
\end{equation}
Figure \ref{decay-and-oscillations} shows the late time decay rate $\Gamma$ and frequency $\omega_0$ of the retarded Green's function, obtained by fitting this ansatz to the numerical solutions. Here we have fixed $J=1$ and $R=2$. As the dissipation strength $K$ is increased, we see a transition from damped oscillations $(\omega_0>0)$ to a purely exponential decay $(\omega_0=0)$ of $G^R(t)$ at around $K\sim 0.4$. This is analogous to the transition observed in the Caldeira-Leggett model, a canonical example of open quantum dynamics \cite{Leggett}. The decay rate $\Gamma$ is expected to increases with stronger dissipation. While this is generally true, in a small window right after the transition, the decay rate decreases as dissipation becomes stronger. This can be interpreted as the quantum Zeno effect in which, frequent measurement or strong environmental coupling (as in this case) can stabilize a quantum state \cite{zeno-experimental, zeno-lindblad1, zeno-lindblad2, zeno-lindblad3}.

\begin{figure}[t]
    \centering
    \begin{subfigure}{0.23 \textwidth}
        \subcaption{$K=0.1$}
        \includegraphics[width=0.8\textwidth]{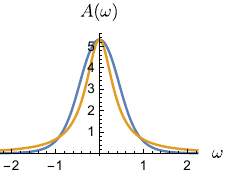}
    \end{subfigure}
    \begin{subfigure}{0.23 \textwidth}
        \subcaption{$K=0.5$}
        \includegraphics[width=0.8\textwidth]{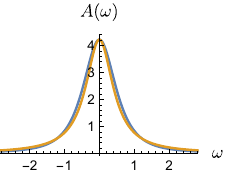}
    \end{subfigure}\\
    \begin{subfigure}{0.23 \textwidth}
        \subcaption{$K=1$}
        \includegraphics[width=0.8\textwidth]{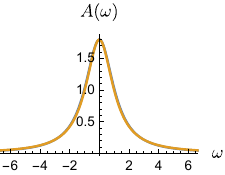}
    \end{subfigure}
    \caption{
    Spectral function (blue)
    compared with Lorentzian (orange) for $J=1, R=2$.}
    \label{spectral-function}
\end{figure}

In the frequency domain, a useful quantity to analyze is the spectral function defined by
\begin{equation}
    A(\omega) = -2\ \text{Im}\,[G^R(\omega)].
\end{equation}
The spectral function can be interpreted as a probability distribution. Indeed, our numerical solutions satisfy the normalization condition $\int_{-\infty}^\infty \frac{d\omega}{2\pi} A(\omega) = 1$. We compare the spectral function to a Lorentzian distribution. Figure \ref{spectral-function} demonstrates that for large dissipation strength $K$, $A(\omega)$ is well approximated by a Lorentzian. We see the same effect at large $R$, that is, for a large number of jump operators. The Lehman representation for Lindbladian systems \cite{Scarlatella_2019} suggests that when the spectral function is Lorentzian the eigenvalue with the largest non-zero real part is purely real. This is consistent with our finite $N$ numerics in the next section.

\subsection{Finite $N$ spectrum}
The stationary Green's functions naturally do not contain all the information about the dynamics of the system. The full information of the dynamics is contained in the eigenvalues and eigenvectors of the Lindbladian \cite{PhysRevA.96.052113}. Here, we will study the eigenvalues, also known as the spectrum, of the Lindbladian \eqref{random-quad-model}. We set $N=10$, which gives a total of $20$ Majorana fields after the doubling described in Section \ref{The operator-state isomorphism and  the Schwinger-Keldysh path integral}. For each set of parameters, we collect 50 realizations of the random Lindbladian to plot the spectrum.

\begin{figure*}[t]
    \centering
    \begin{subfigure}{0.3 \textwidth}
        \includegraphics[width=\textwidth]{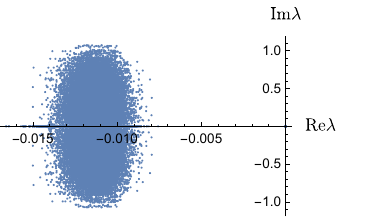}
        \caption{$K=0.1$}
    \end{subfigure}
    \begin{subfigure}{0.3 \textwidth}
        \includegraphics[width=\textwidth]{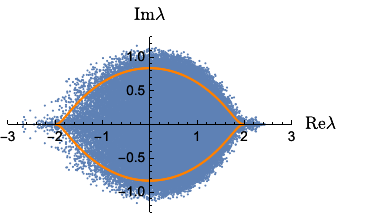}
        \caption{$K=1.5$ (scaled)}
    \end{subfigure}
    \begin{subfigure}{0.3 \textwidth}
        \includegraphics[width=\textwidth]{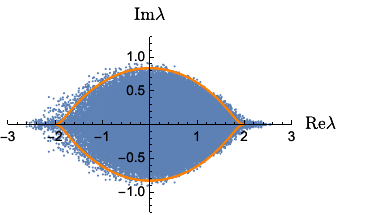}
        \caption{$K=3$ (scaled)}
    \end{subfigure}
    \caption{Spectrum of the Lindbladian \eqref{Random lindblad} for $J=1, R=1$ and varying $K$. (a) The spectrum is elliptical, consistent with the literature. (b) and (c) The spectrum is scaled and shifted according to equation \eqref{scale-shift} before being plotted. As $K$ increases, the boundary of this (scaled) spectrum resembles the lemon-shaped contour derived in \cite{2019PhRvL.123n0403D}.}
    \label{spectrum-vary-K}
\end{figure*}

Figure \ref{spectrum-vary-K} shows the spectra as we vary $K$ while keeping $J=1$ and $R=1$ fixed. For large dissipation strength $K$, the boundary of the spectrum resembles a lemon-shape. We compare this boundary to the spectral boundary of purely disipative fully random Lindblad operators, which was calculated analytically in \cite{2019PhRvL.123n0403D}. To do this comparison, we first scale and shift the eigenvalues as follows:
\begin{equation}
    \lambda_i \rightarrow \sqrt{N} \left( \frac{8}{NK^2}\lambda_i + 1 \right).
    \label{scale-shift}
\end{equation}

Note that the dissipative part of the Lindbladian in Equation \eqref{Random lindblad} contains $\sim N^2$ random entries, whereas a fully random Lindbladian on the $2^{N/2}$-dimensional Hilbert space would contain $\sim 2^{N}$ random entries. Also, only two-body jump operators are considered in this model. Therefore, the boundary of the spectrum may not precisely match the contour derived in \cite{2019PhRvL.123n0403D}, and further investigation is required.
When the dissipation strength $K$ is small relative to the SYK coupling $J$, the spectrum is elliptical. The spectra also show an enhanced density of eigenvalues on the real axis. These features resemble those of random Lindbladians reported in \cite{2019PhRvL.123n0403D, 2019PhRvL.123w4103C, 2020JPhA...53D5303S}. 

\begin{figure*}[t]
    \centering
    \begin{subfigure}{0.3 \textwidth}
        \includegraphics[width=\textwidth]{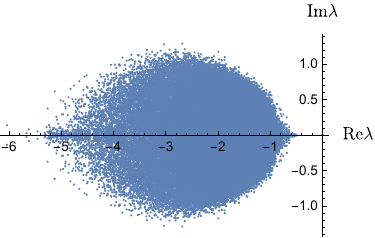}
        \caption{$R=0.5$}
    \end{subfigure}
    \begin{subfigure}{0.3 \textwidth}
        \includegraphics[width=\textwidth]{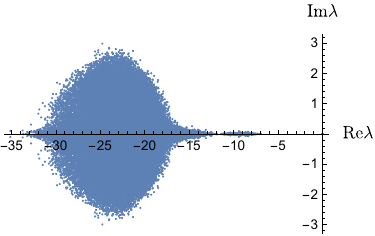}
        \caption{$R=5$}
    \end{subfigure}
    \begin{subfigure}{0.3 \textwidth}
        \includegraphics[width=\textwidth]{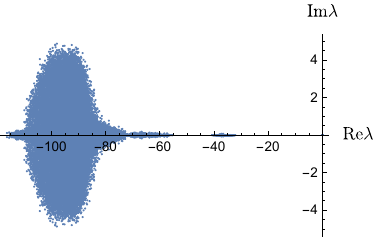}
        \caption{$R=20$}
    \end{subfigure}
    \caption{Spectrum of the Lindbladian \eqref{Random lindblad} for $J=1, K=2$ and varying $R$}
    \label{spectrum-vary-R}
\end{figure*}

The story is quite different when we vary $R$ while keeping $J$ and $K$ fixed. Figure \ref{spectrum-vary-R} shows that the shape of the spectrum changes significantly as we increase $R$. The bulk of the spectrum gets progressively squeezed towards the negative real axis, while clusters of (close to) real eigenvalues are left in its wake.
Similar cluster formation 
and hierarchy of relaxation times
were observed in random Lindbladians
\cite{2019PhRvL.123w4103C,
2020PhRvL.124j0604W, 2021arXiv211013158L}.

In unitary physics, random matrix theory captures the universal features of chaotic dynamics. To understand whether this is also the case in nonunitary physics, it is important to identify physical systems that exhibit nonunitary random matrix behavior. As we have seen, the model \eqref{random-quad-model} indeed serves this purpose.

\section{Summary and outlook}

In this work, we introduced SYK type Lindbladian models and studied their Green's functions in the long time limit, and their spectral properties.
The models admit exact analysis in various limits (large $N$, large $q$, simultaneous large $N$ and $M$ limits). Another merit of the models
is that they exhibit very rich behaviors. In particular, the second model realizes many different behaviors by simply controlling the parameters $J, K,$ and $R$, some of which compare well with different random Lindbladian models studied previously. 
There are many remaining questions. 
We close by listing a few of them. 
First, vast generalizations of the current models are possible, for example, by introducing $p$-body jump operators. Studying wider classes of models would allow us to explore different universal behaviors in open quantum many-body systems. Second, while we studied the distribution of the
eigenvalues of the SYK Lindbladians, a more thorough 
characterization of the spectral properties is necessary. For example, it is of great interest to study the level statistics
\cite{2020PhRvX..10b1019S,
2021arXiv211003444G,
2019PhRvL.123y4101A,
2020PhRvR...2b3286H}.
The level statistics may show an interesting crossover as the distribution crosses over from the lemon shape to the one with many clusters
\cite{2021arXiv211205765P}.
Third, in this work, we mostly focused on stationary properties. However, 
it would be interesting to follow the time evolution by the SYK Lindbladians starting from some initial state. Technically, the Kadanoff-Baym equation can be solved numerically.



{\it Note added:}
Recently, 
\cite{2021arXiv211212109S}
appeared on arXiv,
which has a substantial overlap
with our Section 
\ref{Many Random Jump Operators (Complex coupling)}.

\section*{Acknowledgments}
We thank Kohei Kawabata and Jiachen Li for useful discussions. This work is supported by JST CREST Grant (No.JPMJCR19T3), by the National Science Foundation under Award No.\ DMR-2001181, and by a Simons Investigator Grant from the Simons Foundation (Award No.~566116). This work is supported by the Gordon and Betty Moore Foundation through Grant GBMF8685 toward the Princeton theory program.

\bibliography{bibliography}

\end{document}